\documentclass[12pt]{article}
\usepackage{amsmath,amssymb}
\usepackage{graphicx}
\usepackage[pdftex]{hyperref}
\DeclareGraphicsExtensions{.eps,.bmp,.wmf,.jpg}
\numberwithin{equation}{section}
\def\be{\begin{equation}}
\def\ee{\end{equation}}

\def\bea{\begin{eqnarray}}
\def\eea{\end{eqnarray}}

\title{Holographic dark energy with non-minimal coupling}
\author{L.N. Granda\thanks{ngranda@univalle.edu.co} ,\, and L.D. Escobar\thanks{leon.escobar@correounivalle.edu.co}\\
Department of Physics, Universidad del Valle\\ A.A. 25360, Cali,
Colombia} 

\date{}
\begin{document}
\maketitle

\begin{abstract}
\noindent We study a scalar field non-minimally coupled to the curvature, in the framework of holographic dark energy. We obtain a relation between the  coupling of the scalar field and the holographic DE parameters. In the model without potential we found the EOS parameter in different regions of the parameters, giving rise to accelerated expansion. For some restrictions on the parameters, the model presents quintom behavior.\\ 

\end{abstract}

\section{Introduction}
\noindent 
Recent astrophysical data from distant Ia supernovae observations \cite{SN},\cite{riess} show that the current Universe is not only expanding, it is accelerating due to some kind of  negative-pressure form of matter called dark energy. This dark energy may consist of cosmological constant, conventionally associated with the energy of the vacuum  or alternatively, could came from a dynamical varying scalar field at late times which also account for the missing energy density in the universe. On order to avoid the fine tuning and the coincidence problems, which are connected with the inflationary behavior of the early universe and the late time dark energy dominated regime, the dark energy should have dynamical nature. This stimulates the interest in generalized gravity theories such as the scalar tensor theories of gravity. The scalar fields are allowed from several theories in particle physics and in multidimensional gravity, like Kaluza-Klein theory, String Theory\cite{ST} or Supergravity, in which the scalar field appears in a natural way. The non-minimal coupling normally arises in quantum field theory in curved space time \cite{ford}, \cite{birrel}, and in multidimensional theories like superstring theory \cite{ST}. This models have also been used to study inflationary cosmology \cite{turner}, \cite{amendola1}, \cite{amendola2}, \cite{maeda}, \cite{kasper}.
The late time cosmological solutions using scalar fields has being greatly studied in the last years in the usual model called quintessence \cite{RP}, \cite{wett}, \cite{shani}, \cite{stein}, \cite{copeland97}, \cite{stein1}, \cite{stein2}, where the stability of scaling solutions were studied (i.e. solutions where the scalar energy density scale in the same way like a barotropic fluid) for a field evolving in accordance with an exponential potential and a power law potential, which provide a late time inflation. More exotic approaches to the problem of dark energy using scalar fields are related with K-essence models based on scalar field with non-standard kinetic term \cite{stein3},\cite{chiba}; string theory fundamental scalars known as tachyons \cite{pad}; scalar field with negative kinetic energy, which provides a solution known as phantom dark energy \cite{caldwell}. For a recent review on above mentioned and other approaches to the dark energy problem, see \cite{copeland}.\\
An interesting approach to explain the nature of the dark energy is based on some facts of quantum gravity known as holographic principle \cite{thooft,susskind,bousso}. This principle establishes a connection between the short distance (ultraviolet) cut-off  and the long distance (infrared) cut-off, given by a restriction on the size of the system in such a manner that prevents the formation of black holes with size larger than the size of the system \cite{cohen, hsu}. Applied to the dark energy issue, if we take the whole universe into account, then the vacuum energy related to this
holographic principle is viewed as dark energy, usually called holographic dark energy. Based on this principle, the proposal of holographic dark energy have been developed in \cite{cohen, hsu, li}. Different IR cutoff scales such as Hubble scale,
particle horizon and event horizon have been proposed to establish the holographic dark energy models. The Hubble scale can not give rise to an accelerated universe \cite{hsu}, while the event horizon can produce an accelerated expansion \cite{li}. In this work we consider a scalar field non minimally coupled to the curvature, with a new ingredient of holographic density as proposed in \cite{granda}, as a source of dark energy and analyze whether the accelerating universe takes place.  The holographic dark energy in the presence of non-minimal coupling have been presented in \cite{ito,setare},  and the cosmological implications of modified gravity non minimally coupled with matter, have been considered in \cite{odintsov1}.

\section{Non-minimal coupling and holography}
Here we assume the generalization of the holographic principle in the presence of non-minimally coupled scalar field. Let us start with the following scalar field action 

\be\label{eq1}
S_{\phi}=\int d^{4}x\sqrt{-g}\left[\frac{1}{2\kappa^2}R-\frac{1}{2}\partial_{\mu}\phi\partial^{\mu}\phi-\frac{1}{2}\xi R{\phi}^{2}\right].
\ee
where $\kappa^2=8\pi G$ and we are considering the spatially-flat Friedmann–Robertson–Walker (FRW) metric with signature$ (-,+,+,+)$. Variation with respect to the metric, and assuming that the scalar field $\phi$ has only time dependence, gives the following modified Friedman Eqs. in the flat FRW background \cite{uzan,chiba1}
\be\label{eq2}
  H^2 = \frac{\kappa^2}{3} \left(\frac{1}{2}\dot{\phi}^2+6\xi H\phi\dot{\phi}+3\xi H^2\phi^2+\rho_{\Lambda}\right)
\ee
which corresponds to the $(00)$ component of the variation with respect to the metric, and for the $(11)$ component it is obtained
\be\label{eq3}
  -2\dot{H}-3H^2=\kappa^2\left[(\frac{1}{2}-2\xi)\dot{\phi}^2-2\xi(\phi\ddot{\phi}+2H\phi\dot{\phi})-\xi(2\dot{H}+3H^2)\phi^2+p_{\Lambda}\right]
\ee
where $H$ is the Hubble parameter, and $\rho_{\Lambda}$, $p_{\Lambda}$ are the energy density and pressure of the holographic dark energy. The equation of motion of the scalar field is the modified Klein-Gordon equation
\be\label{eq4}
  \ddot{\phi}+3H \dot{\phi}+6\xi \left( \dot{H} + 2H^2 \right)
  \phi=0 
\ee
Although the above equations may include the potential, in this case we show that the effect of accelerated expansion can be obtained without the need to introduce a potential.
Due to the non-minimal coupling from \ref{eq2} or \ref{eq3}, the effective gravitational coupling can be expressed as
\be\label{eq5}
\tilde{G}=\frac{G}{1-8\pi \xi G\phi^2}
\ee
where $G$ is the constant Newtonian coupling. The relative time variation of the gravitational coupling obtained from \ref{eq5} leads to the current value of $\dot{\tilde{G}}/\tilde{G}$
\be\label{eq5a}
\dot{\tilde{G}}/\tilde{G}=-\frac{6\xi^2\kappa^2\phi_0^2}{(1-6\xi)(1-\xi\kappa^2\phi_0^2)}
\ee
With the effective gravitational coupling (\ref{eq5}), the Eqs. (\ref{eq2}) and (\ref{eq3}) become respectively
\be\label{eq6}
H^2=\frac{1}{3}\frac{\kappa^2}{1-\xi\kappa^2\phi^2}\left(\frac{1}{2}\dot{\phi}^2+6\xi H\phi\dot{\phi}+\rho_{\Lambda}\right)
\ee
and 
\be\label{eq7}
-2\dot{H}-3H^2=\frac{\kappa^2}{1-\xi\kappa^2\phi^2}\left[(\frac{1}{2}-2\xi)\dot{\phi}^2-2\phi\ddot{\phi}-4\xi H\phi\dot{\phi}+p_{\Lambda}\right]
\ee
from which we can read the modified density and pressure of the scalar field:
\be\label{eq8}
\tilde{\rho_{\phi}}=\frac{1}{2}\dot{\phi}^2+6\xi H\phi\dot{\phi}
\ee
and
\be\label{eq9}
\tilde{p_{\phi}}=(\frac{1}{2}-2\xi)\dot{\phi}^2-2\xi\phi\ddot{\phi}-4\xi H\phi\dot{\phi}
\ee
Taking the infrared cut-off as given in \cite{granda} with the redefined gravitational coupling given by Eq. (\ref{eq5}), leads to the holographic dark energy density in the presence of non-minimally coupled scalar field 
\be\label{eq10}
\rho_{\Lambda}=\frac{3}{\kappa^2}\left(1-\xi\kappa^2\phi^2\right)\left(\alpha H^2+\beta\dot{H}\right)
\ee
Looking for power law solutions of this model, we take the time dependence of $a$ and $\phi$ of the form
\be\label{eq11}
a(t)=a_0 t^p,\,\,\,\,\,\,\,\, \phi=\phi_0t^s
\ee
From the Eq. of motion \ref{eq4} it follows the relation between the powers $p$ and $s$
\be\label{eq12}
s(s-1)+3ps+6\xi p(2p-1)=0
\ee
replacing Eq. \ref{eq11} in Eq. \ref{eq2} we obtain the additional two relations involving the coupling constant and the holographic parameters
\be\label{eq13}
p\left(1-\alpha\right)+\beta=0,\,\,\,\,\,\,\,\, s=-12\xi p
\ee
from equations (\ref{eq12}) and (\ref{eq13}) we can express $\xi, p$ and $s$ in terms of the holographic parameters
\be\label{eq14}
\xi=\frac{1-\alpha+4\beta}{24\beta},\,\,\,\,\, p=\frac{1}{4-24\xi}=\frac{\beta}{\alpha-1},\,\,\,\,\,\, s=\frac{1-\alpha+4\beta}{2(1-\alpha)}
\ee
the condition for accelerated expansion translates into the inequality
\be\label{eq15}
1-\alpha+\beta>0
\ee
which in terms of the coupling $\xi$, means $1/8<\xi<1/6$. Note (from (\ref{eq14})) that if $\xi>1/6$, $p$ becomes negative, giving rise to phantom behavior. For the case of $\alpha<1, \beta>0$, $\xi$ is always positive and $p$ negative, reproducing also phantom behavior. From the second of Eqs. (\ref{eq14}) it follows that negative values of $\xi$ give $p<1/4$, which give rise to decelerated expansion, and therefore for the proposed solution, negative $\xi$ are not interesting.\\
Replacing $\phi$ and $a$ in Eq. \ref{eq10} for the holographic density, it follows
\be\label{eq16}
\rho_{\Lambda}=\frac{3}{\kappa^2}p(\alpha p-\beta)\left(t^{-2}-\xi \kappa^2\phi_0^2t^{2s-2}\right)
\ee
and applying the conservation equation
\be\label{eq17}
\dot{\rho_{\Lambda}}+3H\left(\rho_{\Lambda}+p_{\Lambda}\right)=0
\ee
which follows from Eqs. \ref{eq2} and \ref{eq3} combined with the Bianchi identity, we obtain the pressure density $p_{\Lambda}$ which defines the equation of state 
\be\label{eq18}
w_{\Lambda}=-1+\frac{8}{3}\left(1-6\xi\right)-\frac{8\xi^2\kappa^2\phi_0^2}{t^{24\xi p}-\xi\kappa^2\phi_0^2}
\ee
replacing $\xi$ and $p$ from \ref{eq14} and changing to the redshift variable it is obtained
\be\label{eq19}
w_{\Lambda}=-1+\frac{2}{3}\frac{\alpha-1}{\beta}-\frac{(1-\alpha+4\beta)^2}{72\beta^2}\frac{\kappa^2\phi_0^2}{(1+z)^{-\frac{1-\alpha+4\beta}{\beta}}-\xi\kappa^2\phi_0^2}
\ee
From Eqs. \ref{eq6} and \ref{eq7} clearly follows that the total or effective EOS is constant
\be\label{eq20}
w_{eff}=\frac{\tilde{p_{\phi}}+p_{\Lambda}}{\tilde{\rho_{\phi}}+\rho_{\Lambda}}=-1-\frac{2\dot{H}}{3H^2}=-1+\frac{2}{3}\frac{\alpha-1}{\beta}
\ee
as is expected for the power-law behavior. This expression coincides with the limit $\kappa^2\phi_0^2<<1$ from (\ref{eq19}). An additional relation between $\alpha$, $\beta$ and $\xi$ can be obtained by constraining the current ($z=0$) EOS of the holographic dark energy (i.e. $w_{\Lambda0}\approx-1$), and analyzing the behavior of the EOS $w(z)$ in the vicinity of the conformal coupling. First we can define a critical value for $\alpha$ (and therefore for $\xi$), using the condition that at present time (z=0) the EOS parameter takes the value $w_{\Lambda0}=-1$. Given the values of $\kappa^2\phi_0^2=1$ and $\beta=0.5$, from Eqs. (\ref{eq18},\ref{eq19}) follows that $w_{\Lambda}(z=0)=-1$ at $\alpha_c=1.1655$ or equivalently $\xi_c=1/6-0.014\approx 0.1527$. For values of $\alpha>\alpha_c$, for instance for $\alpha=1.23$ as shown in Fig. 1, the present holographic EOS parameter is greater than $-1$, showing quintessence behavior and maintaining this behavior in the future; and for values below the $\alpha_c$, for instance $\alpha=1.1$ the present value of $w_{\Lambda}$ shows phantom behavior but changes to quintessence in the future. Fig. 1 shows the holographic EOS for this three cases. In all this cases we preserved the inequality $\xi\kappa^2\phi^2<1$ in order to maintain the effective gravitational constant positive. Similar studies have been done in \cite{ito} for the holographic non minimal coupling with the Hubble scale as infrared cut-off, and in \cite{setare} for a variety of non-minimally coupled scalar fields.
\begin{center}
\includegraphics[scale=0.9]{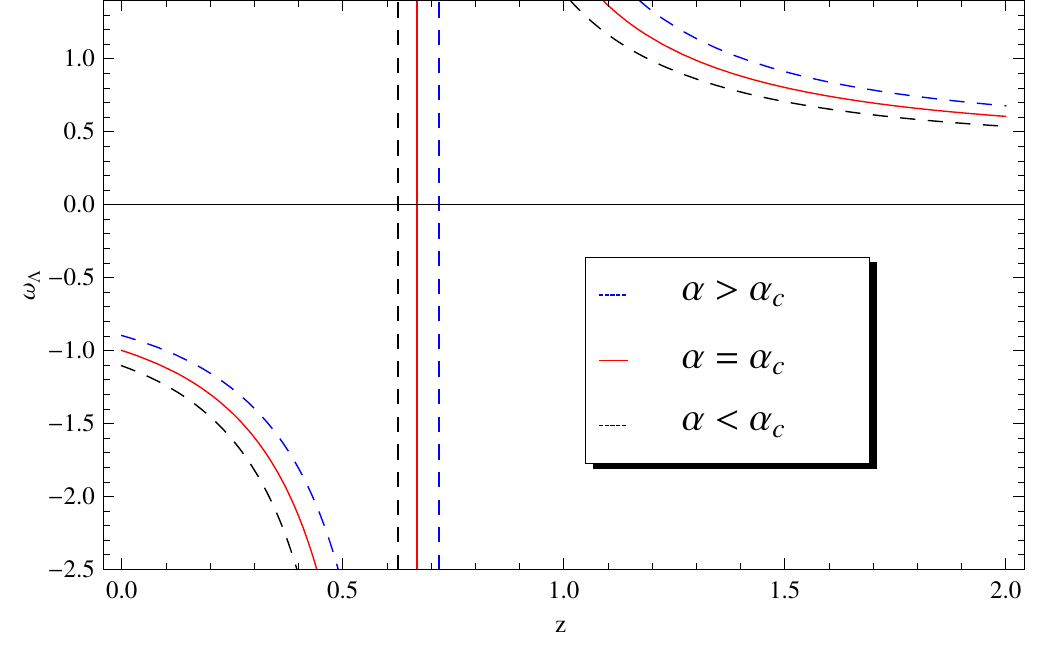}
\end{center}
{\it The behavior of the holographic EOS parameter as function of the redshift, for fixed values of $\kappa^2\phi^2=1$, $\beta=0.5$ and three different cases corresponding to the present quintessence, cosmological constant and phantom behavior. All these states evolve in the future to quintessence}\\
By changing $\kappa^2\phi^2$ and $\beta$ we can define the corresponding new $\alpha_c$ and $\xi_c$. If we take $\kappa^2\phi^2<<1$, then from Eq. (\ref{eq19}) follows that $\alpha_c\rightarrow1$ or $\xi_c\rightarrow1/6$ and the divergence in Eq. (\ref{eq19}) moves to the high redshift region (distant past). In this case the quintessence phase moves to $\alpha>1$ and the phantom phase to $\alpha<1$. 
Note that from Eq. \ref{eq6}, in order to satisfy the current observational constraints on the relative time variation of the gravitational constant $\left|\dot{G}/G\right|$ when $\xi\rightarrow1/6$, the magnitude of $\kappa^2\phi^2$ should be of the order $\kappa^2\phi^2\sim6(1-6\xi)$. In the case of Fig. 1 this variation is of the order $\left|\dot{G}/G\right|\sim 1/t_0\sim6\times10^{-11} yr^{-1}$, which is between the limits set by some observations \cite{uzan1}. 

\section{Discussion}
The proposed holographic dark energy model \cite{granda}, in the presence of non-minimally coupled scalar field yields accelerated expansion, and from the present to the future evolution is able to show quintom behavior, under some viable restrictions on the parameters. We analyzed the different allowable restrictions on the model parameters giving rise to the present acceptable values of the dark energy EOS parameter.
Note that the divergence at $z\approx 0.6-0.7$ in Fig. 1, comes from the denominator in Eq. (\ref{eq19}) and can be moved to the higher redshift region for values of $\kappa^2\phi_0^2<<1$, or to the future for $\kappa^2\phi_0^2>>1$ (in this last case we should preserve the inequality $\xi\kappa^2\phi^2<1$ in order to prevent $\tilde{G}$ being negative, according to Eq. \ref{eq5}; and this means that $\xi\rightarrow0$ which is not interesting because gives rise to decelerated expansion (\ref{eq14})). It's worth to mention that for this specific selection of the parameters, this divergence occurs at redshift close to the redshift transition ($z_T\sim 0.5-0.7$ according to the observations)  and separates the regions of decelerated (right) and accelerated (left) expansion as shown in Fig. 1. The divergence in the EOS can be removed by introducing a dark matter term of the form $\Omega_{m0}(a+z)^3$, and defining the new EOS in the form $w=p_{\Lambda}/(\rho_{\Lambda}+\rho_m)$. This guarantees the presence of a term (1+z) to some positive power in the denominator, which prevents the denominator to have zeros and smooths the transition deceleration-acceleration. 

Focusing on the future, this model is free of divergences (under the condition $\kappa^2\phi_0^2<1$) and  the current EOS ($z=0$) is able to describe quintessence, cosmological constant and phantom behavior, but under the considered power-law  solutions (\ref{eq11}) all states evolve to quintessence in the future. In this manner, the model describes quintom behavior without introducing phantom scalars.
Of course, those are the possibilities between the limitations of the proposed power-law solution, and more general and free of singularities solutions, are currently under study.

\section*{Acknowledgments}
This work was supported by the Universidad del Valle.

\end{document}